\documentclass[12pt]{article}

\catcode`\@=11
\@addtoreset{equation}{section}

\global\arraycolsep=2pt
\usepackage{amsbsy,amssymb,latexsym,amsfonts,amsmath,cite}
\usepackage{graphicx,color}

\allowdisplaybreaks

\newcommand\e{\mathrm{e}}

\newcommand\nn{\nonumber}

\newcommand{\sch}{Schr\"odinger }

\begin{document}
\begin{flushright}
\parbox{4.2cm}
{UCB-PTH-08/77 \\
OIQP-08-14 \\
NSF-KITP-08-141}
\end{flushright}

\vspace*{0.5cm}

\begin{center}
{\Large \bf 
Interacting SUSY-singlet matter 

in non-relativistic Chern-Simons theory
}
\vspace*{1cm}\\
{Yu Nakayama$^{\flat}$,
Makoto Sakaguchi$^{\sharp}$ and Kentaroh Yoshida$^{\natural}$
}
\end{center}
\vspace*{-0.2cm}
\begin{center}
$^{\flat}${\it Berkeley Center for Theoretical Physics, \\ 
University of California, Berkeley, CA 94720, USA
} 
\vspace*{0.5cm}\\
$^{\sharp}${\it Okayama Institute for Quantum Physics\\
1-9-1 Kyoyama, Okayama 700-0015, Japan} 
\vspace*{0.5cm}\\
$^{\natural}${\it Kavli Institute for Theoretical Physics, \\ 
University of California, Santa Barbara, CA 93106, USA} 
\vspace{0.5cm}\\
{\tt nakayama@berkeley.edu} \qquad 
{\tt makoto\_sakaguchi@pref.okayama.jp} \qquad 
{\tt kyoshida@kitp.ucsb.edu} 
\end{center}

\vspace{0.8cm}

\begin{abstract} 
We construct an example of supersymmetric Chern-Simons-matter theory
with a matter field transforming as a singlet representation of the
supersymmetry algebra, where the bosonic and fermionic degrees of
freedom
do not match. This is obtained as a non-relativistic limit of the
$\mathcal{N}=2$ Chern-Simons-matter theory in 1+2 dimensions, where the
particle and anti-particle coexist. We also study the index to investigate the mimatch of bosonic and fermionic degrees of freedom.
\end{abstract}

\thispagestyle{empty} 

\setcounter{page}{0}

\newpage

\section{Introduction} 

The Coleman-Mandula theorem \cite{Coleman:1967ad} and its supersymmetric
extension \cite{Haag:1974qh} play a significant role in classifying {\it
relativistic} supersymmetry (SUSY) algebras and their
representations. The basic dynamical SUSY algebra
\begin{eqnarray}
\{ Q, Q^*\} = 2H
\end{eqnarray}
demands that all the relativistic fields should transform as a
non-trivial representation of the SUSY algebra. If a field $\phi$ were a
singlet under the SUSY, i.e. $[Q,\phi] = 0$ and $[Q^*,\phi]=0$,
then $\phi$ would be a singlet under the Hamiltonian time-evolution
$[H,\phi] = 0$ from the super Jacobi identity. Thus $\phi$ could not be a
dynamical field.

This argument may be modified in the {\it non-relativistic} (NR) system,
where there is no direct analogue of the Coleman-Mandula theorem. Hence
there is a theoretical possibility to realize
a theory containing a dynamical field that transforms as a singlet under
the SUSY transformation.

Indeed, the Galilean algebra can admit a supersymmetric extension with
the so-called kinematical SUSY:
\begin{eqnarray}
\{Q, Q^*\} = 2M\,,
\end{eqnarray}
where $M$ is a mass operator, or more generally an internal
symmetry. Then, there is no no-go theorem that forbids a singlet field
under the kinematical SUSY charges $Q$ and $Q^*$. Explicit field
theoretical models that possess such a kinematical SUSY algebra alone
can be found in the study of NR Chern-Simons system (e.\ g.\ in
\cite{NSRY,NSY})\footnote{Super Schr\"odinger algebras with
dynamical SUSY in 1+2 dimensions are discussed in
\cite{Leblanc:1992wu,Sakaguchi:2008rx,Sakaguchi:2008ku}. 
Super \sch algebras containing only the kinematical SUSY are 
also presented \cite{Sakaguchi:2008rx,Sakaguchi:2008ku} and 
the algebras of this type may be related to the gravity dual 
discussed in \cite{Hartnoll:2008rs}.}.

Still, it is quite a non-trivial challenge to construct an example of
supersymmetric field theory with a singlet representation under this
algebra. One of the reasons is that we do not have a superfield
formulation for NR system. Without the superfield formation, the SUSY transformation is not independent of the action and both of them should be determined at once, especially
when a gauge field is introduced.

In this letter, instead of following the standard manner, we present a
theory with a SUSY-singlet matter by taking a NR limit of the
relativistic $\mathcal{N}$=2 Chern-Simons-matter theory
\cite{Leblanc:1992wu}. Along the line of discussions in \cite{NSRY}, the 
relativistic Chern-Simons-matter theory is indeed a ubiquitous generating source 
of many inequivalent (super) Schr\"odinger invariant field theories. The exotic theory 
with a SUSY-singlet field, which we have just mentioned and we will pursue here, 
is also contained in the resulting theories. 

Since the bosonic and fermionic degrees of freedom do not match in our SUSY-singlet field theories while preserving supersymmetry, it would be interesting to study the index-like
object introduced in \cite{Nakayama:2008qm}. We will conclude this letter by computing the index for primary operators of these SUSY-singlet NR Chern-Simons theories. 

\section{Relativistic $\mathcal{N}=2$ Chern-Simons-matter theory}

The relativistic $\mathcal{N}$=2 Chern-Simons-matter theory in 1+2
dimensions was originally constructed in \cite{Leblanc:1992wu}. The
action is composed of the Chern-Simons term $S_{\rm CS}$ and matter part
$S_{\rm M}$ as follows\footnote{We use the same spinor convention as in
\cite{NSRY}.}:
\begin{eqnarray}
S_{\rm rel} &=& S_{\rm CS} + S_{\rm M} ~, \nn
\\ 
S_{\rm CS} &=& \int dt d^2x\, \frac{\kappa}{4 } 
\epsilon^{\mu\nu\lambda} 
A_{\mu} F_{\nu\lambda} = \int dt d^2x \left[
\kappa A_0 F_{12} + \frac{\kappa}{2c}\epsilon^{ij}
\partial_t A_i A_j \right] \qquad (i,j=1,2) \,,  \nn \\ 
S_{\rm M}
&=& \int dt d^2x \biggl[- (D_{\mu}\phi)^{\ast} D^{\mu}\phi 
- i \bar{\psi} \gamma^{\mu}D_{\mu}\psi \nn \\  
&&  \qquad \quad -\left(
\frac{ e^2}{ \kappa c^2} 
\right)^2 
|\phi |^2 
\left(|\phi|^2 - v^2 \right)^2 
+ \frac{e^2}{\kappa c^2} \left( 3|\phi|^2 -v^2\right)
i\bar{\psi}\psi \biggl] \,. \label{LLW}
\end{eqnarray}
The mass $m$ is read off as $m^2c^2=(\frac{e^2}{\kappa c^2})^2v^4$.
The action (\ref{LLW}) is invariant under the following SUSY
transformation
\begin{eqnarray}
\delta A_{\mu} &=& 
\frac{e}{\kappa c}\bar{\alpha}\gamma_{\mu} 
\psi\phi^{\ast} 
+\frac{e}{\kappa c}\bar{\psi}\gamma_{\mu}\alpha\phi 
~,\label{rels}  \nn \\ 
\delta \phi &=& -i\bar{\alpha} \psi\,, \\ \nn
\delta \psi &=&  -\gamma^{\mu}\alpha D_{\mu}\phi 
+ \frac{e^2}{\kappa c^2} \alpha \phi (v^2-|\phi|^2)\,.  
\end{eqnarray}
Here $\alpha$ is a 2-component complex Grassmann variable and hence
(\ref{LLW}) has $\mathcal{N}$=2 supersymmetries in 1+2 dimensions.

\section{Non-relativistic limit and singlet SUSY}

In this section, we study NR limits of the Chern-Simons-matter
theory. Since the Chern-Simons part does not show any change in NR
limits, the non-trivial difference only appears in the matter sector.

In order to study NR limits of (\ref{LLW}), first of all, let us
expand the fields as follows:
\begin{eqnarray}
\phi &=& \frac{1}{\sqrt{2m}} \left[\,\e^{-im c^2 t} \Phi
+\e^{im c^2 t} \hat{\Phi}^{\ast} \right]\,, \nn \\ 
\psi &=& \sqrt{c} \left[\,\e^{-i m c^2 t} \Psi
+ \e^{i m c^2 t} C \hat{\Psi}^{\ast} \right]\,, \nn 
\end{eqnarray}
where $\Psi = (\Psi_1, \Psi_2)^t$ and $\hat{\Psi} = (\hat{\Psi}_1, \hat{\Psi}_2)^t$ are two component complex Grassmann-valued fields, and $C=i\sigma_2$ is a charge conjugation matrix. Here, ``hat''
denotes the anti-particle.

\medskip 

The naive NR limit ($c\to \infty$) with keeping both particle and
anti-particle leads to the following matter action\footnote{The absolute
square of fermions is defined as $|\Psi|^2=\Psi^* \Psi$\,.}
\begin{eqnarray}
S &=& \int\!dt d^2x\,\biggl[
i\Phi^{\ast}D_t\Phi + i\hat{\Phi}^{\ast}\hat{D}_t\hat{\Phi} 
-\frac{1}{2m}\left[(D_i\Phi)^{\ast}D_i\Phi 
+ (\hat{D}_i\hat{\Phi})^{\ast}\hat{D}_i\hat{\Phi}\right] \nn \\ 
&& \hspace*{1.5cm} + i\Psi_1^{\ast}D_t\Psi_1 
+ i\hat{\Psi}_1^{\ast}\hat{D}_t\hat{\Psi}_1  
-\frac{1}{2m}\left[(D_i\Psi_1)^{\ast}D_i\Psi_1  
+ (\hat{D}_i\hat{\Psi}_1)^{\ast}\hat{D}_i\hat{\Psi}_1\right] \nn \\ 
&& \hspace*{1.5cm} -\frac{e}{2mc}F_{12}(|\Psi_1|^2 
- |\hat{\Psi}_1|^2) + \lambda (|\Phi|^2 + |\hat{\Phi}|^2)^2 
+ 2\lambda |\Phi|^2|\hat{\Phi}|^2
\nn  \\ 
&& \hspace*{1.5cm} + 3\lambda (|\Phi|^2 + |\hat{\Phi}|^2)
(|\Psi_1|^2 + |\hat{\Psi}_1|^2) \biggr]\,, \label{naive}
\end{eqnarray}
where $\Psi_2$ and $\hat{\Psi}_2$ have been removed by using the
equations of motion,
\begin{eqnarray}
\Psi_2 = - \frac{1}{2mc}D_+\Psi_1 + \mathcal{O}(1/c^2)\,, \qquad 
\hat{\Psi}_2 = \frac{1}{2mc}\hat{D}_+\hat{\Psi}_1 + \mathcal{O}(1/c^2)\,. 
\end{eqnarray}
We have also introduced the coupling constant $\lambda
=\frac{e^2}{2mc\kappa}$. It is easy to see that the action is invariant
under the bosonic Schr\"odinger symmetry \cite{Sch,JP}.

\medskip 

The SUSY transformation at the leading order is given by 
\begin{eqnarray}
&& \delta_1 \Phi = -\sqrt{2mc}\,\alpha^{(1)\ast}\Psi_1\,, \qquad 
\delta_1 \hat{\Phi} = \sqrt{2mc}\,\alpha^{(2)}\hat{\Psi}_1\,, \nn \\ 
&& \delta_1 \Psi_1 = \sqrt{2mc}\,\alpha^{(1)}\Phi\,, \qquad 
\delta_1\hat{\Psi}_1 = -\sqrt{2mc}\,\alpha^{(2)\ast}\hat{\Phi}\,, \nn \\ 
&& \delta_1 A_0 =\frac{e}{\sqrt{2mc}\kappa}\left[
\alpha^{(1)\ast}\Psi_1\Phi^{\ast} -\alpha^{(1)} \Psi_1^{\ast}\Phi 
-\alpha^{(2)\ast}\hat{\Psi}_1^{\ast}\hat{\Phi} 
+\alpha^{(2)}\hat{\Psi}_1\hat{\Phi}^{\ast}
\right]\,,    \nn \\ 
&& \delta_1 A_i =0\,.
\end{eqnarray} 
The SUSY transformation at the next-to-leading order is 
\begin{eqnarray}
&& \delta_2 \Phi = -\frac{1}{\sqrt{2mc}}\,\alpha^{(2)\ast}D_+\Psi_1\,, 
\qquad \delta_2 \hat{\Phi} =
\frac{1}{\sqrt{2mc}}\,\alpha^{(1)}\hat{D}_+\hat{\Psi}_1\,, \nn \\ 
&& \delta_2 \Psi_1 = -\frac{1}{\sqrt{2mc}}\, \alpha^{(2)}D_-\Phi \,, 
\qquad \delta_2 \hat{\Psi}_1 = \frac{1}{\sqrt{2mc}}\, \alpha^{(1)\ast}
\hat{D}_-\hat\Phi\,, \nn \\ 
&& \delta_2 A_0 = \frac{e}{(2mc)^{3/2}\kappa}\biggl[
-\alpha^{(2)\ast}(D_+\Psi_1)\Phi^{\ast} +
\alpha^{(2)}(D_+\Psi_1)^{\ast}\Phi \nn \\ 
&& \hspace*{4cm} +\alpha^{(1)\ast}(\hat{D}_+\hat{\Psi}_1)^{\ast}\hat{\Phi} 
-\alpha^{(1)}(\hat{D}_+\hat{\Psi}_1)\hat{\Phi}^{\ast} 
\biggr] \,, \nn \\ 
&& \delta_2 A_+ = \frac{2ie}{\sqrt{2mc}\kappa}\left[
\alpha^{(2)}\Psi_1^{\ast}\Phi +\alpha^{(1)\ast}\hat{\Psi}_1^{\ast}\hat{\Phi}
\right]\,, \nn \\ 
&& \delta_2 A_- =\frac{2ie}{\sqrt{2mc}\kappa}\left[
\alpha^{(2)\ast}\Psi_1\Phi^{\ast} +\alpha^{(1)}\hat{\Psi}_1\hat{\Phi}^{\ast}
\right]\,. 
\end{eqnarray}
These transformations are directly obtained from the NR limit of \eqref{rels}.
Note that the SUSY parameters $\alpha^{(1)}$ and $\alpha^{(2)}$ are not
separated.

\medskip 

The naive NR action $S_\mathrm{CS}+S$ with (\ref{naive}), however, is not invariant even under the
SUSY transformation at the leading order. We have no clear understanding
of this result, but it is possible to improve the situation by adding a
four-fermi interaction without spoiling any symmetries of the naive
action. The improved action is given by
\begin{eqnarray}
S &=& \int\! dtd^2x\,\biggl[
i\Phi^{\ast}D_t\Phi + i\hat{\Phi}^{\ast}\hat{D}_t\hat{\Phi} 
-\frac{1}{2m}\left[(D_i\Phi)^{\ast}D_i\Phi 
+ (\hat{D}_i\hat{\Phi})^{\ast}\hat{D}_i\hat{\Phi}\right] \nn \\ 
&& \hspace*{1.5cm} + i\Psi_1^{\ast}D_t\Psi_1 
+ i\hat{\Psi}_1^{\ast}\hat{D}_t\hat{\Psi}_1  
-\frac{1}{2m}\left[(D_i\Psi_1)^{\ast}D_i\Psi_1  
+ (\hat{D}_i\hat{\Psi}_1)^{\ast}\hat{D}_i\hat{\Psi}_1\right] \nn \\ 
&& \hspace*{1.5cm} -\frac{e}{2mc}F_{12}(|\Psi_1|^2 
- |\hat{\Psi}_1|^2) + \lambda (|\Phi|^2 + |\hat{\Phi}|^2)^2 
+ 2\lambda |\Phi|^2|\hat{\Phi}|^2
\nn  \\ 
&& \hspace*{1.5cm} + 3\lambda (|\Phi|^2 + |\hat{\Phi}|^2)
(|\Psi_1|^2 + |\hat{\Psi}_1|^2) + 2\lambda |\Psi_1|^2|\hat{\Psi}_1|^2
\biggr]\,. \label{improve}
\end{eqnarray}
This improved model has 4 supersymmetries at the leading order and there
is no SUSY transformation at the next-to-leading order\footnote{The
next-to-leading SUSY will be resurrected when we only keep particle
degrees of freedom as in \cite{Leblanc:1992wu}. See the discussion in
the subsequent subsections.}. This is in accordance with the general
expectation from \cite{NSRY}: since the SUSY transformation is not
separated from the leading order and the next-to-leading order, the
second SUSY is not realized.  The improved action now gives a new super
\sch invariant field theory with 4 (real) supercharges. The SUSY charges
\begin{eqnarray}
Q_1 = \sqrt{2m} \int\! d^2x\, \Psi_1^*\Phi\,, \qquad 
Q_2 = -\sqrt{2m}  \int\! d^2x\, \hat{\Psi}_1^* \hat{\Phi}\,,
\end{eqnarray}
satisfy the following anti-commutation relations
\begin{eqnarray}
\{Q_1, Q_1^*\} = 2m N_1\,, \qquad 
\{Q_2, Q_2^*\} = 2m N_2\,, 
\end{eqnarray}
where we have introduced the particle number density $N_1$ and the
anti-particle number density $N_2$ defined as, respectively, 
\begin{eqnarray}
N_1 = \int\! d^2x\, \left( |\Phi|^2 + |\Psi_1|^2 \right)\,, \qquad 
N_2 = \int\! d^2x\, \left( |\hat{\Phi}|^2 + |\hat{\Psi}_1|^2\right)\,.
\end{eqnarray}

\subsection{Consistent truncation}

There are several different ways to take NR limits by reducing the
matter contents. We will consider below:
\begin{center}
\begin{enumerate}
\item ~~all particle case: PP \quad  ($\hat{\Phi}=\hat{\Psi}$=0) 
\item ~~a singlet SUSY 1: BP \quad ($\hat{\Psi}$ = 0)
\item ~~a single SUSY 2: PB \quad ($\hat{\Phi}$ = 0)
\end{enumerate}
\end{center}
The sequence of alphabets denotes the degrees of freedom we hold in the
NR limit: particle (P), anti-particle (A), both particle and
anti-particle (B), and neither of them (N). Since the NR limit preserves
the particle number density and the anti-particle number density
separately for both bosons and fermions, all the truncations discussed
here are consistent (in the sense of strong condition introduced in
\cite{NSRY}).

The first case is nothing but \cite{Leblanc:1992wu}, so we will not
discuss it here except for pointing out the fact that the superconformal
symmetry emerges in contrast to the full limit presented above. This is
related to the emergence of the dynamical SUSY that is lacking in the
full NR action.

Here, we concentrate on more exotic possibilities such as BP or PB to
construct a supersymmetric field theory with a singlet representation
under the SUSY. It is clear that the other possibilities (up to exchange
of particles with anti-particles) PA, BN, NB, PN and NP lead to
non-supersymmetric theories. We emphasize the ubiquity of the
relativistic Chern-Simons-matter theory to give birth to many
inequivalent (super) Schr\"odinger invariant field theories.

\subsection{Singlet SUSY 1}

We construct a singlet supersymmetric field theory from the BP case. With our
ansatz ($\hat{\Psi} = 0$), the matter action reads
\begin{eqnarray}
S &=& \int\!dt d^2x\,\biggl[
i\Phi^{\ast}D_t\Phi + i\hat{\Phi}^{\ast}\hat{D}_t\hat{\Phi} 
-\frac{1}{2m}\left[(D_i\Phi)^{\ast}D_i\Phi 
+ (\hat{D}_i\hat{\Phi})^{\ast}\hat{D}_i\hat{\Phi}\right] \nn \\ 
&& \hspace*{1.5cm} + i\Psi_1^{\ast}D_t\Psi_1  
-\frac{1}{2m}(D_i\Psi_1)^{\ast}D_i\Psi_1
-\frac{e}{2mc}F_{12}|\Psi_1|^2  \nn \\ 
&& \hspace*{1.5cm} + 
\lambda (|\Phi|^2 + |\hat{\Phi}|^2)^2 
+ 2\lambda |\Phi|^2|\hat{\Phi}|^2 
+ 3\lambda (|\Phi|^2 + |\hat{\Phi}|^2)|\Psi_1|^2
\biggr]\,. \label{improve1}
\end{eqnarray}
Note that this action (\ref{improve1}) has no subtlety associated with the
four-fermi term because it vanishes identically due to the ansatz.

The NR action $S_\mathrm{CS}+S$ with (\ref{improve1})
 is invariant under the following SUSY transformation
\begin{eqnarray}
&& \delta_1 \Phi = -\sqrt{2mc}\,\alpha^{(1)\ast}\Psi_1\,, \qquad 
\delta_1 \hat{\Phi} = 0\,, \nn \\ 
&& \delta_1 \Psi_1 = \sqrt{2mc}\,\alpha^{(1)}\Phi\,, \nn \\ 
&& \delta_1 A_0 =\frac{e}{\sqrt{2mc}\kappa}\left[
\alpha^{(1)\ast}\Psi_1\Phi^{\ast} -\alpha^{(1)} \Psi_1^{\ast}\Phi 
\right]\,,    \nn \\ 
&& \delta_1 A_i =0\,.
\end{eqnarray}
It is easy to check explicitly that the next-to-leading SUSY
transformation (dynamical SUSY) is broken.

It is not difficult to see that the action is invariant under the
Schr\"odinger symmetry, so the NR limit here yields a Schr\"odinger
invariant field theory with 2 real supercharges. The anti-commutator of
the SUSY charges
\begin{eqnarray}
Q = \sqrt{2m}\int d^2x \, \Psi_1^* \Phi
\end{eqnarray}
 can be computed as
\begin{eqnarray}
\{Q,Q^*\} = 2m N_1 \ ,
\end{eqnarray}
where $N_1$ is the conserved charge associated with the number density
of particles\footnote{In addition, the model has two $U(1)$ symmetries
associated with the total mass operator and the fermion number.}:
\begin{eqnarray}
N_1 = \int d^2x \left( |\Phi|^2 + |\Psi_1|^2\right) \ .
\end{eqnarray}

The bosonic field $\hat{\Phi}$ transforms as a singlet under the SUSY
transformation, and it does not have its fermionic partner. Nevertheless
the field $\hat{\Phi}$ non-trivially interacts with other fields.

It would be possible to consider various supersymmetric 
deformations of the theory.
First of all, the SUSY does not fix the coefficient in front of
$|\hat{\Phi}|^4$, which is a SUSY-singlet, and such a deformation
corresponds to a classical marginal deformation of the super
Schr\"odinger invariant theory. 
Secondly,
we can change the
coefficient of 
\[
\lambda_1 |\Phi|^2|\hat{\Phi}|^2 + \lambda_2
|\Psi|^2|\hat{\Phi}|^2
\] 
as long as we demand 
\[
\lambda_1 = \lambda_2 +\frac{e^2}{2mc\kappa}
\]
with electric charge $e^2/\kappa$ fixed.

\subsection{Singlet SUSY 2}

Similarly, we can construct a singlet supersymmetric field theory from
the PB case. With our ansatz ($\hat{\Phi} = 0$), the matter action reads
\begin{eqnarray}
S &=& \int\!dt d^2x\,\biggl[
i\Phi^{\ast}D_t\Phi 
-\frac{1}{2m}(D_i\Phi)^{\ast}D_i\Phi  +  i\Psi_1^{\ast}D_t\Psi_1 
+ i\hat{\Psi}_1^{\ast}\hat{D}_t\hat{\Psi}_1  \nn \\ 
&& \hspace*{1.5cm} 
-\frac{1}{2m}\left[(D_i\Psi_1)^{\ast}D_i\Psi_1  
+ (\hat{D}_i\hat{\Psi}_1)^{\ast}\hat{D}_i\hat{\Psi}_1\right] 
-\frac{e}{2mc}F_{12}(|\Psi_1|^2- |\hat{\Psi}_1|^2)  \nn \\ 
&& \hspace*{1.5cm}  
+ \lambda |\Phi|^4  + 3\lambda |\Phi|^2
(|\Psi_1|^2 + |\hat{\Psi}_1|^2) + 2\lambda |\Psi_1|^2|\hat{\Psi}_1|^2
\biggr]\,. \label{improve3}
\end{eqnarray}
Here, we have used improved action and added the four-fermi term.

The NR action $S_\mathrm{CS}+S$ with (\ref{improve3}) is invariant under the following SUSY transformation
\begin{eqnarray}
&& \delta_1 \Phi = -\sqrt{2mc}\,\alpha^{(1)\ast}\Psi_1 \,, \nn \\ 
&& \delta_1 \Psi_1 = \sqrt{2mc}\,\alpha^{(1)}\Phi\,,  \qquad 
\delta_1 \hat{\Psi}_1 = 0\,, \nn \\ 
&& \delta_1 A_0 =\frac{e}{\sqrt{2mc}\kappa}\left[
\alpha^{(1)\ast}\Psi_1\Phi^{\ast} -\alpha^{(1)} \Psi_1^{\ast}\Phi 
\right]\,,    \nn \\ 
&& \delta_1 A_i =0\,.
\end{eqnarray}
It is an easy task to check that the next-to-leading SUSY transformation
(dynamical SUSY) is broken.

It is not difficult to see that the action is invariant under the
Schr\"odinger symmetry, so the NR limit here yields a Schr\"odinger
invariant field theory with 2 real supercharges. The anti-commutator of
the SUSY charges
\begin{eqnarray}
Q = \sqrt{2m}\int d^2x \Psi_1^* \Phi
\end{eqnarray}
 can be computed as
\begin{eqnarray}
\{Q,Q^*\} = 2m N_1 \ ,
\end{eqnarray}
where $N_1$ is the conserved charge associated with the number density
of particles\footnote{In addition, the model has two $U(1)$ symmetries
associated with the total mass operator and the fermion number.}:
\begin{eqnarray}
N_1 = \int d^2x \left( |\Phi|^2 + |\Psi_1|^2 \right) \ .
\end{eqnarray}

The fermionic field $\hat{\Psi}_1$ transforms as a singlet under the
SUSY transformation, and it does not have its bosonic
partner. Nevertheless the field $\hat{\Psi}_1$ non-trivially interacts
with other fields.

The model is essentially obtained by replacing
$\hat{\Phi}$ by $\hat{\Psi}_1$ in the BP theory because $\hat{\Phi}$ is
a singlet representation and it can simply be replaced with 
another singlet field $\hat{\Psi}_1$. The additional Pauli interaction,
which does not exist in the BP theory, is also a SUSY-singlet, so there
is no problem here.

More generally, if a SUSY-singlet piece $F(Q_i)$ could be constructed out of fields $Q_i$ with non-trivial representations of SUSY, it would be possible to couple it to a SUSY-singlet field $S$ in a SUSY invariant Lagrangian as
\[
\delta L = f(S, \partial S) \times
F(Q_i)\,.
\]
As is discussed in the introduction, for
dynamical SUSY (as in the relativistic case), there is no such a candidate 
of $F(Q_i)$. This is equivalent to the well-known fact that the
relativistic supersymmetric field theory does not have a SUSY invariant
Lagrangian, but the invariance is always only up to total derivative
terms.

The kinematical SUSY allows nontrivial $F(Q_i)$ as we have explicitly seen in this section. The form of $f(S,\partial S)$ (as well as
$F(Q_i)$) is partially fixed by the classical Schr\"odinger
invariance. It would be interesting to see whether the quantum
Schr\"odinger invariance makes the parameters of the theory more
restrictive.\footnote{For example, the $\mathcal{N}=2$ PP limit
\cite{Leblanc:1992wu} has vanishing beta functions. See \cite{NSRY} and
references therein.}

\section{Index for primary operators}
Our BP theory and PB theory do not have balanced bosonic degrees of freedom and fermionic degrees of freedom. Because of this mismatch, we may suspect that the virtue of SUSY (namely, the Bose-Fermi cancellation) might be lost.  From the 
anti-commutation relation, however, the potential trouble could appear
only in zero $N_1$ sector. In addition, the non-zero $N_1$ sector may be 
taken as a superselection sector due to the particle number
conservation in the NR system. With this regard, we would like to study the index-like
object $\mathrm{Tr} (-1)^F e^{-\beta N_1}$ in this section. 

The NR supersymmetric conformal field theories discussed in the previous section have a non-trivial involutive anti-automorphism (see \cite{Nakayama:2008qm} for details). We can use it to define the index as 
\begin{eqnarray}
I(x) = \mathrm{Tr} (-1)^F e^{-\beta N_1} x^D \ ,
\end{eqnarray}
where $2mN_1 = \{Q,Q^*\}$ for our BP, PB theories. This index counts the operators annihilated by $Q^*$, and we will see that the index does not depend on $\beta$. In order to distinguish operators that contribute to the index, we have introduced the chemical potential $x$ that couples to the dilatation $D$. Note that $D$ commutes with $Q$ and $Q^*$ so that the Bose-Fermi cancellation is intact.

Because of the complicated singular-vector structure of the zero particle number  ($M=0$) sector in the representation theory of Schr\"odinger group, we only study the index for primary operators annihilated both by the Galilean boost $G_i$ and the special conformal transformation $K$. In the $e\to 0 $ limit, the computation of the index boils down to the counting of gauge invariant operators, which may be obtained by the integration over the $U(1)$ holonomy \cite{Nakayama:2008qm} as 
\begin{eqnarray}
I_p(x) \equiv \mathrm{Tr}_{\mathrm{primary}} (-1)^F e^{-\beta N_1} x^D =  \int_0^{2\pi} \frac{d\theta}{2\pi} \exp \left[\sum_i\sum_{n=1}^{\infty} \frac{1}{n} f_i(n\theta, n\beta, x^n)\right] \ , 
\end{eqnarray}
where the summation $i$ is over the species of primary fields and the corresponding single particle indices $f_i$ are shown in table 1.

\begin{table}[tb]
\begin{center}
\begin{tabular}{c|c|c|c|c|c|c|c|c}
 Letters        & $\Phi$& $\Phi^*$ & $\hat{\Phi}$ & $\hat{\Phi}^* $ & $\Psi_1$ & $\Psi^*_1$ & $\hat{\Psi}_1$ & $\hat{\Psi}^*_1$ \\
 \hline
 $N_1$&     $ +1 $& $-1$ & $0$ &$0$ & $+1$ & $-1$ &$0$&$0$ \\
  $M$&     $+1$  & $-1$ & $+1$ &$-1$ & $+1$ &$-1$ &$+1$ &$-1$ \\
$D$&     $1 $   &  $1$ & $1$ & $1$ & $1$ & $1$ &$1$ &$1$  \\
$f(\theta, \beta,x)$ & $x e^{-\beta}e^{i\theta}$ &$x e^{\beta}e^{-i\theta}$&$ x e^{i\theta}$&$ x e^{-i\theta}$ & $-x e^{-\beta}e^{i\theta}$&$-x e^{\beta}e^{-i\theta}$ &$-x e^{i\theta}$&$-x e^{-i\theta}$ \\ 
 \end{tabular}
\end{center}
\caption{List of the letters contributing to the index for primary operators.}
\label{tab:1}
\end{table}%

The direct integration gives the index for primary operators as
\begin{eqnarray}
I_p(x;\mathrm{BP}) &=& \frac{1}{1-x^2} \cr
I_p(x;\mathrm{PB}) &=& 1-x^2 \ .
\end{eqnarray}
The index is independent of $\beta$ due to the Bose-Fermi cancellation in the non-zero $N_1$ sector as advocated, but the non-zero index shows a mismatch between the bosonic and fermionic degrees of freedom in the 
zero $N_1$ sector.

\section*{Acknowledgements}

We would like to thank S.~Ryu for fruitful discussions and early
collaboration of this project. The work of NY was supported in part by
the National Science Foundation under Grant No.\ PHY05-55662 and the UC
Berkeley Center for Theoretical Physics. The work of MS was supported in
part by the Grant-in-Aid for Scientific Research (19540324) from the
Ministry of Education, Science and Culture, Japan. The work of KY was
supported in part by the National Science Foundation under Grant No.\
PHY05-51164 and JSPS Postdoctoral Fellowships for Research Abroad.


\begin{thebibliography}{99}
\bibitem{Coleman:1967ad}
  S.~R.~Coleman and J.~Mandula,
  ``ALL POSSIBLE SYMMETRIES OF THE S MATRIX,''
  Phys.\ Rev.\  {\bf 159} (1967) 1251.

\bibitem{Haag:1974qh}
  R.~Haag, J.~T.~Lopuszanski and M.~Sohnius,
  ``All Possible Generators Of Supersymmetries Of The S Matrix,''
  Nucl.\ Phys.\  B {\bf 88} (1975) 257.


\bibitem{NSRY}
 Y.~Nakayama, S.~Ryu, M.~Sakaguchi and K.~Yoshida,
  ``A family of super Schr\"odinger invariant Chern-Simons matter systems,''
  arXiv:0811.2461 [hep-th].

\bibitem{NSY}
Y.~Nakayama, M.~Sakaguchi and K.~Yoshida, 
``Non-Relativistic M2-brane Gauge Theory and New Superconformal Algebra,'' 
in preparation. 

\bibitem{Leblanc:1992wu}
  M.~Leblanc, G.~Lozano and H.~Min,
  ``Extended superconformal Galilean symmetry in Chern-Simons matter systems,''
  Annals Phys.\  {\bf 219} (1992) 328
  [arXiv:hep-th/9206039].

\bibitem{Sakaguchi:2008rx}
  M.~Sakaguchi and K.~Yoshida,
  ``Super Schr\"odinger algebra in AdS/CFT,'' \\
  J.~Math.~Phys. {\bf 49} (2008) 102302
  [arXiv:0805.2661 [hep-th]].

\bibitem{Sakaguchi:2008ku}
  M.~Sakaguchi and K.~Yoshida,
  ``More super Schr\"odinger algebras from psu(2,2$|$4),''
  JHEP {\bf 0808} (2008) 049 
  [arXiv:0806.3612 [hep-th]].

\bibitem{Hartnoll:2008rs}
  S.~A.~Hartnoll and K.~Yoshida,
  ``Families of IIB duals for nonrelativistic CFTs,''
  arXiv:0810.0298 [hep-th].
\bibitem{Nakayama:2008qm}
  Y.~Nakayama,
  ``Index for Non-relativistic Superconformal Field Theories,''
  JHEP {\bf 0810} (2008) 083
  [arXiv:0807.3344 [hep-th]].
\bibitem{Sch}
  C.~R.~Hagen,
  ``Scale and conformal transformations in Galilean-covariant field theory,''
  Phys.\ Rev.\  D {\bf 5} (1972) 377. \\ 
U.~Niederer, 
``The maximal kinematical invariance group
of the free Schr\"odinger equation,'' 
Helv.\ Phys.\ Acta {\bf 45} (1972) 802.

\bibitem{JP}
 R.~Jackiw and S.~Y.~Pi ,
 ``Soliton Solutions to the gauged nonlinear Schr\"odinger  
equation on the plane,'' 
Phys.\ Rev.\ Lett.\  {\bf 64} (1990) 2969: \\
  ``Classical and quantal nonrelativistic Chern-Simons theory,''
  Phys.\ Rev.\  D {\bf 42} (1990) 3500
  [Erratum-ibid.\  D {\bf 48} (1993) 3929].  


\end{thebibliography}
\end{document}